# FIRST PRINCIPLES STUDY OF THE SI(557)-AU SURFACE

**Daniel Sánchez-Portal**[1] , **Richard M. Martin**[2]

[1]*Centro Mixto CSIC-UPV/EHU and Donostia International Physics Center (DIPC), Pº Manuel de Lardizabal 4, E-20018 Donostia, Spain.*
[2]*Department of Physics and Materials Research Laboratory, University of Illinois, Urbana, Illinois 61801, US*

**Abstract**

We have performed a density functional study of fifteen different structural models of the Si(557)-Au surface reconstruction. Here we present a brief summary of the main structural trends obtained for the more favourable models, focusing afterwards in a detailed description of the atomic structure, electronic properties and simulated STM images of the most stable model predicted by our calculations. This structure is in very good agreement with that recently proposed from X-ray diffraction measurements by Robinson *et al*. [Phys. Rev. Lett. 88, (2002) 096194].

The stepped Si(577)-Au reconstruction is formed after the deposition of ~0.2 monolayers of gold on top of vicinal Si(111) surfaces, as described in Ref. [1]. The steps go along the [1-10] direction, and each terrace, with a width of ~19 Å, is supposed to contain one row of gold atoms running parallel to the step edge. With such structure, one could naively expect that this system would exhibit only a single one-dimensional (i.e. only dispersing along the direction parallel to the chains) metallic band, originated in the 6s states of Au. However, the photoemission experiments of Segovia *et al*. [1] revealed two peaks in the vicinity of the Fermi level that were interpreted in terms of the spin-charge separation in a Luttinger liquid [1]. More recently, new photoemission experiments [2] as well as density functional (DFT) calculations [3], pointed to the possibility that the two observed features could instead correspond to two different quasiparticle bands, crossing the Fermi level at different, although neighboring, points of the Brillouin zone. However, the interpretation given to the appearance of these two bands was very different in both works. While in the theoretical study [3] these bands were assigned to the Si atoms neighboring to the Au chain, Losio *et al.* [2] correlated their appearance with the observation, using Scanning Tunnelling Microscopy (STM), of two prominent chain-like features running parallel to each step. They concluded then that two Au chains, rather than one, should decorate each step of the surface, with a metallic band associated with each of them.

The possibilities of solving this and other puzzles posed by the photoemission[1,2] and STM [2] data of this system are limited by the many unknowns still remaining about the main characteristics of the Si(557)-Au surface. A crucial example is the detailed atomic structure of the reconstruction, for which only very recently some reliable experimental information has become available from the X-ray diffraction experiments by Robinson *et al.* [4]. The knowledge of the structure is however instrumental since, in the absence of accurate models, it is meaningless to attempt a detailed comparison between the observed and calculated excitation spectra. Therefore it becomes necessary to perform an exhaustive structural investigation in order to find likely candidates for the atomic reconstruction. In this paper we present a step along this direction. We have applied first-principles density functional (DFT) methodology to the study of the atomic and electronic structure of fifteen different structural models of the Si(557)-Au surface. Here we will focus on the presentation of the results for our most stable model, which we propose as a good candidate for the structure of this surface reconstruction, leaving for a future publication a more detailed account of our study [5].

Our calculations have been performed using the SIESTA method [6,7], which allows for standard DFT calculations for systems containing hundreds of atoms. We have used here the local approximation to DFT [8,9], Troullier-Martins norm-conserving pseudopotentials [10], and a basis set of numerical atomic orbitals obtained from the solution of the atomic pseudopotentials at slightly excited energies [6,7,11,12][1]. A Brillouin zone sampling of 10 inequivalent k-points, and a real-space grid equivalent to a plane-wave cutoff of 100 Ry (up to 40 k-points and 200 Ry in convergence tests) has been employed. This guarantees the convergence of the total energy, for a given basis set, within ~20 meV/Au (~0.5 meV per atom in the supercell).

The steps of the Si(557)-Au surface have been modelled by slabs containing three or four silicon double layers. The Si atoms in the bottom layer are saturated with hydrogen, and remain at the bulk ideal positions during the relaxation process. The supercells contain then up to 116 atoms. Due to this large number of atoms we have used a double-$\zeta$ (DZ) basis for the structural optimisations, which include two different functions (i.e. two distinct radial shapes) to represent the 3s states of Si, and another two for the 3p shell, and the 6s and 5d shell of Au. Once the equilibrium geometry has been obtained, a more complete basis set (DZP) including polarization orbitals (i.e. a d shell for Si, and a p shell for Au) has been used for the calculation of the electronic band structure and the STM images. This larger basis set has also been utilized for the structural relaxation in those systems containing less than 60 atoms. In all these cases the structures and relative surface energies found with DZ and DZP basis are very similar. The results obtained with slabs of three or four Si bilayers provide very similar results, indicating that the results are well converged, and here we only present results calculated with the thicker slabs. To avoid artificial stresses, the lattice parameter parallel to the surface is fixed at the calculated bulk value, 5.49 and 5.41 Å respectively using DZ and DZP basis (to be compared with the experimental value of 5.43 Å).

Our results show that it is energetically favourable for the Au atoms to substitute some of the Si atoms in the surface layer. For example, sitting the metal atoms in such substitutional positions in the middle of the terrace lowers the surface energy[2] by ~1 eV/Au compared to configurations where Au sites on top of the terraces, saturating some of the existing dangling bonds. Furthermore, the Au substitution in the terraces is at least 0.5 eV/Au more stable than the adsorption of the Au atoms at the step edge. This is quite striking since the step edges are frequently considered as the natural sites to attach extra atoms, where they can more efficiently create bonds with the substrate and contribute to release the stress in this highly strain regions. However, the tendency of the metal atoms to avoid the step edge in the present system is clear: not only the adsorption sites at the step edge are unfavourable, but even those substitutional sites neighboring to the step edge are ~0.3 eV/Au less favourable that those located well within the terraces. This clearly points out to the importance of the configuration adopted by the step edge for the stability of the Si(557)-Au reconstruction. One interesting consequence of the preference for substitutional positions is that, in spite of the large Au-Au distances, the rigidity of the Si structure keeps the wire stable against dimerization. This allows the Au chain to remain metallic on the surface. This was confirmed by our calculations for simple structural models, presented in Ref.[3], where *two* one-dimensional metallic bands appeared associated with each wire in the surface.

---

[1]We have used an 'energy shift' of 200 meV. The corresponding radii are 5.26, 6.43 and 6.43 a.u. for the s, p and d states states of Si, 6.24, 6.24 and 4.5 in the case of Au, and 5.08 for the s state of H.

In our search for more stable structural models we have tried to reduce the number of dangling bonds in the surface incorporating Si adatoms. The addition of one row of adatoms on the step terrace is energetically favourable in at least one case, with an energy gain of ~0.14 eV/Au. For some structural models it is also possible to include a second row of adatoms, saturating then most of the dangling bonds in the surface. However, this addition always results in an increase of the surface energy. This additional row of adatoms can only be accommodated on the terrace if the Au atoms site in the step edge, otherwise it must be directly located over the step edge. In both cases the rebonding of the Si atoms at the step edge is largely inhibited, pointing again to a direct link between the stability of the proposed model and the structure of the step edge. Fig. 1 (a) shows the structure of our most stable model, out of the fifteen studied here. This structure is in very good agreement with the recent proposal by Robinson *et al.*[4] from the analysis of X-ray diffraction measurements. The main ingredients of the reconstruction are: a row of adatoms sited in the terrace, but neighboring to the edge of the next step; the chain of Au atoms sited in the middle of the step's terrace; the atoms at, and close to, the step edge are strongly rebonded. Gold atoms seem to accommodate in the structure without much strain, with a calculated Au-$Si_1$ bond length (see atom labelling

---

[2] To compare the energies of systems containing different numbers of Si atoms, the Si chemical potential has

in Fig.1 (a)) of 2.45 Å and 2.43 Å for the Au-$Si_2$ distance, quite close to our theoretical bulk Si-Si distance of 2.375 Å. This is again in agreement with the experimental proposal of Ref.[4]. The presence of adatoms produces a doubling of the periodicity along the step. In spite of this, the dimerization along the Au chain is still negligible (less than 0.01 Å). A more significant dimerization is observed for the $Si_2$-$Si_2$ distances, which alternate between 3.91 and 3.78 Å. This "dimerization" can also be regarded as an alternating $Si_2$-Au-$Si_2$ bond angle of 101.8° and 109.6°. The lateral positions of the atoms in the steps edge are fixed by symmetry, however, the adatoms induce an appreciable buckling, $e_2$ atoms (see Fig.1 (a)) lying 0.43 Å below $e_1$ atoms. We will see below that some of these distortions have visible consequences in the band structure and STM images. Of special interest is the structure adopted by the step edge, reminiscent of the building block of the Si(111)-Ag-(3x1) reconstruction [13,14], which we identify as a key element in the stability of this structure. The atoms in the step edge move forward to saturate the dangling bonds in the neighboring terrace. This movement is favored by the tendency of the surface layer to expand, with a change from a $sp^3$ to $sp^2$ hybridization. In analogy to the case of the Si(111)-Ag-(3x1) surface [14], we proposed that this is accompanied by the creation of a complex bonding pattern that can be described by a Si double bond (2.35 Å), marked as *db* in Fig. 1(a), which is further stabilized by the presence of the Si atom immediately below. This atom cannot be considered as strictly bound to any of the atoms forming the double bond due to the large distances (2.57 and 2.83 Å), but hybridises with the π "molecular orbital" form by them. The calculated energy difference between the structures which present and not present this characteristic rebonding is tipically of ~0.5 eV/Au. This is close to the 0.52 eV energy difference between the extended Pandey model of the (3x1) reconstruction, which does not present a Si double

---

been taken equal to the total energy of the bulk silicon at the equilibrium lattice parameter.

bond, and the most stable one containing the Si double bond, giving further support to our identification [14].

The calculated band structure of this model along some of the high symmetry lines of the two dimensional Brillouin zone can be found in Fig(1) (b), while the constant current STM images simulated using Tersoff-Hamman theory [15] are shown in Fig. 2. Γ-K-M line is parallel to the steps, while M-Γ is perpendicular to them. The different surface bands and resonances have been marked according to their main atomic character, which we identify with the help of a Mulliken population analysis [16]. Two surface bands are marked with solid lines: an unoccupied band ~0.5 eV above the Fermi level ($E_F$) which comes from the adatoms (labelled in Fig.1 (a) as *ad*), while the occupied one ~0.2 eV below $E_F$ is originated by the rest atoms (label *rest*). The topology and energy position of these bands, as well as the relaxed positions of the adatoms and rest atoms, are similar to those obtained, for example, for the Si(111)-(2x2) adatom reconstruction [17]. The bands marked with open squares correspond to the atoms located in the step edge: the fully occupied has a larger weight in the $e_2$ atoms, while the band pinning $E_F$ is mainly due to $e_1$ atoms. In Fig. 2 it becomes evident that the most relevant features visible in the STM images are originated in the surface bands commented to this point (adatoms, rest atoms and, step edge) and, therefore, do not contain almost any information about the position and bonding configuration of the Au atoms. The Si adatoms dominate the image when probing empty states above 0.5 eV, while the rest atoms and the step edge become more visible at negative voltages (filled states), as expected. The buckling of the step edge is clearly appreciable in the STM topography, with the lower ($e_2$) atoms better imaged at negative polarities. According to this, we interpret the two chain-like features observed with STM in Ref.[2] as due to the row of adatoms and the step edge, respectively. The band marked with open triangles in Fig.1 (b) corresponds to the π* antibonding combination of the states of the Si atoms forming the double bond, being equivalent to the $S^{-1}$ band of Ref.[14]. The sole surface bands with some Au character appeared in Fig. 1 (b) marked using circles. However, these bands are mainly due to the Si atoms neighboring to the Au chain, and only gain some gold character through hybridization, since the 6s state of Au is completely filled and lies well below $E_F$ consistently with the larger electron affinity of gold. The very flat occupied bands marked with open circles correspond to the $Si_1$ atoms. While the dispersive bands marked with solid circles are associated with the $Si_2$ atoms. The opening of a gap above $E_F$, induced by the above commented alternation of two different $Si_2$-Au-$Si_2$ bond angles, is clearly visible for these bands close to the Brillouin zone boundary.

In conclusion, we have performed a density functional study of several structural models of the Si(557)-Au surface. The most stable model predicted by our calculations is in good agreement with the structure recently proposed by Robinson *et al.* [4], giving further support to their model of the reconstruction. We have presented a detailed analysis of the electronic structure of the surface, and its correlation with the stability of the model. Our simulated STM images seem to be in agreement with the experimental results, indicating that most prominent features observed with STM are related not related with the Au chains, but with the silicon structures on the surface

**Acknowledgements**

We thank F. J. Himpsel and I. K. Robinson for many illuminating discussions, and NCSA for computational resources. D.S.P acknowledges support from the Spanish Ministerio de Ciencia y Tecnología and CSIC under the "Ramón y Cajal" program.

**Figure Captions**

**Figure 1**

(a**)** Relaxed atomic positions of the most stable model for the Si(557)-Au surface. See the text for the labelling of the different atoms. (b) Electronic band structure of the surface reconstruction presented in panel (a). Energies are referred to the Fermi level, and surface states have been marked with different symbols according to their main atomic character (see the text).

**Figure 2**

Simulated STM images for the surface model presented in Fig.1 (a) obtained using Tersoff-Hamann theory. The bias voltage of each image is indicated, with positive (negative) bias corresponding to empty (filled) states of the sample. The position of the rows of adatoms and Au atoms, as well as the step edges, are schematically indicated. The top panels show a lateral view of the images obtained with +1.8 V (left) and −1.8 V (right) giving a better idea of the topography obtained with STM, and of the registry with the atomic positions

**Figure 1**

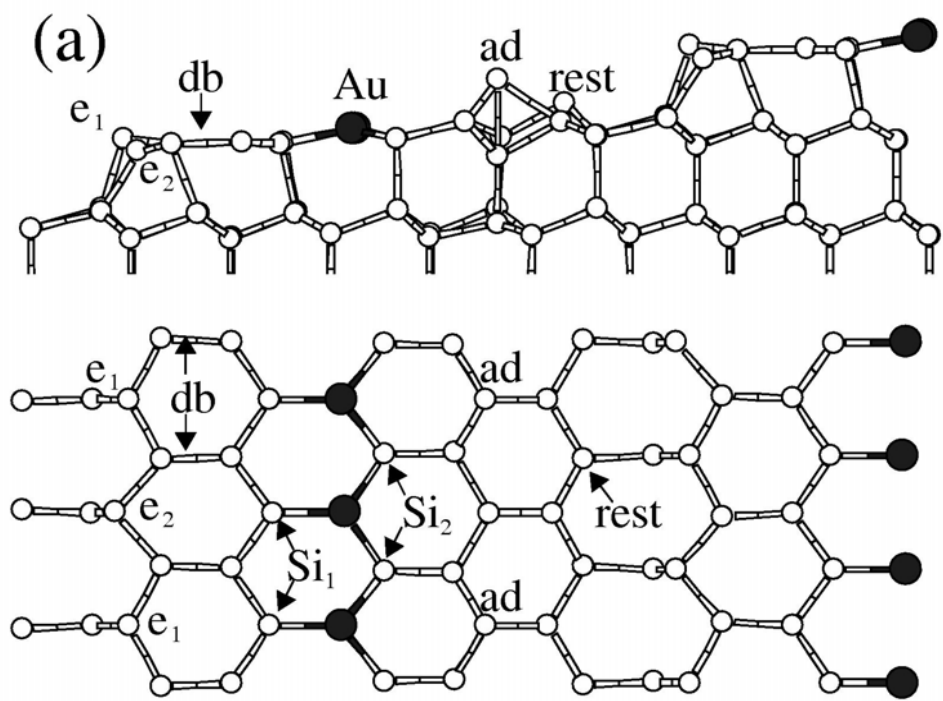

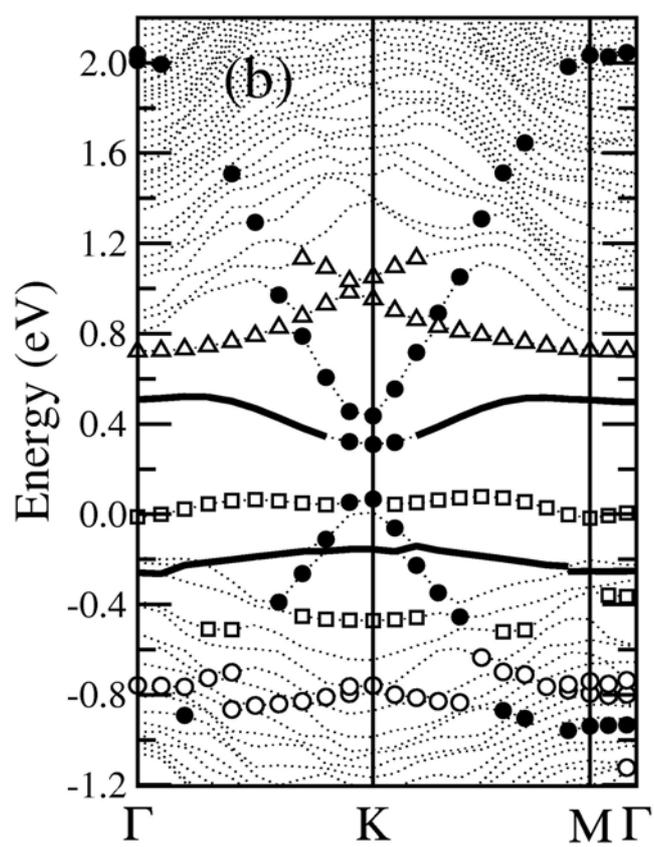

**Figure 2**

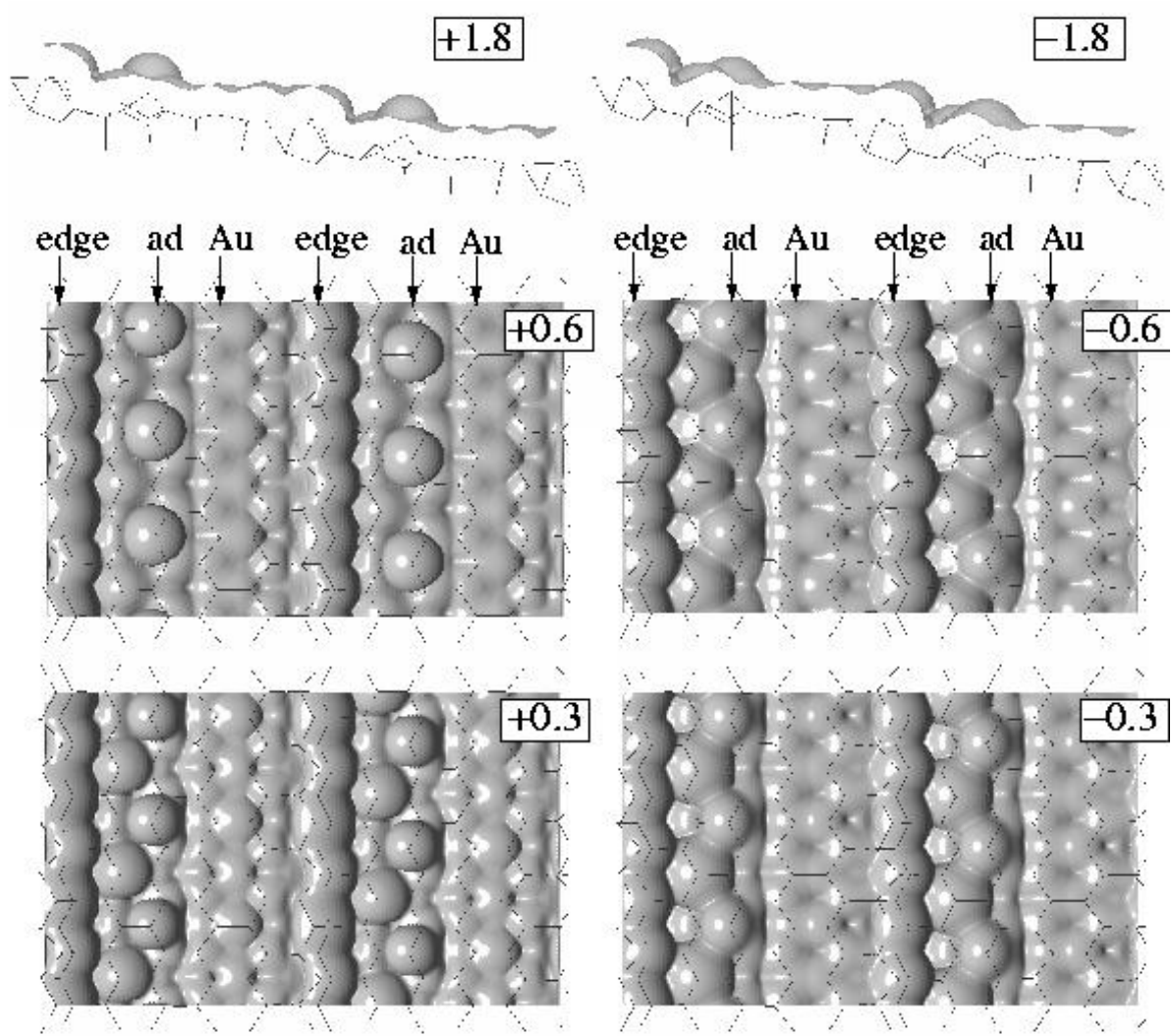